\documentclass{PoS}
\usepackage{amssymb,amsmath, amsthm}
\usepackage{wrapfig}
\usepackage{epsf,graphics,graphicx,epsfig}
\usepackage[usenames,dvipsnames,svgnames,table]{xcolor}
\def\lsim{\raise0.3ex\hbox{$<$\kern-0.75em\raise-1.1ex\hbox{$\sim$}}}
\def\gsim{\raise0.3ex\hbox{$>$\kern-0.75em\raise-1.1ex\hbox{$\sim$}}}

\title{The last word(s) on CPOD 2013}

\ShortTitle{The last word(s) on CPOD 2013}

\author{Frithjof Karsch\thanks{This work has been supported in part 
by contract DE-AC02-98CH10886 with the U.S. Department of Energy.}\\
        Physics Department, Brookhaven National Laboratory, Upton, NY 11973, USA\\
Fakult\"at f\"ur Physik, Universit\"at Bielefeld, D-33615 Bielefeld,
Germany
\\
        E-mail: \email{karsch@bnl.gov}}

\abstract{Higher order moments of net conserved charge fluctuations, 
in particular net baryon number
and net electric charge, are sensitive thermodynamic observables that respond
strongly to critical behavior in strong interaction matter. In order to use them also as
a sensible probe to detect critical behavior in heavy ion experiments we not only need
to better understand 
the relation between chemical freeze-out in heavy ion collision and the
QCD phase boundary, we also need to verify that charge 
fluctuations measured experimentally indeed correspond to thermal conditions
as described by equilibrium QCD. This requires a model independent 
characterization of thermal conditions for which cumulants of conserved charge 
fluctuations themselves are ideally suited.
} 

\FullConference{8th International Workshop on Critical Point and Onset of Deconfinement,\\
		March 11 to 15, 2013\\
		Napa, California, USA}

\begin{document}

\section{Introduction}
A major incentive for the heavy ion research programs at RHIC, the SPS and
LHC as well as for the planning of the future facilities FAIR and NICA is
the prospect to gain insight into the supposably rich phase structure 
of strongly interacting matter at high temperature and non-zero net
baryon number density. Beam energy scan (BES) programs  
at RHIC and the SPS are performed to search for evidence for a second order 
phase transition point, the critical endpoint, which has been postulated 
to exist at high temperature and non-zero net baryon number density or, 
equivalently, non-zero baryon chemical potential ($\mu_B$) \cite{CEP}. 

Experiments at RHIC and LHC provide plenty of evidence
that locally equilibrated hot and dense matter, the
quark-gluon plasma, is formed 
in ultra-relativistic heavy ion collisions. Thermal photons emitted from 
the early phase of the expanding fireball \cite{photons}
suggest that the temperature was 
at least twice as large as the QCD transition temperature, $T_c\simeq 155$~MeV
\cite{WBTc,hotQCDTc}.
The analysis of elliptic flow of different particle species provides 
evidence that the observed collective behavior arises from the interaction 
of partonic degrees of freedom \cite{flow}. Their dominance, however,
gets lost at low beam energies. The BES at the SPS 
provided first evidence for this so-called\footnote{The start for a 
systematic search for the Onset of Deconfinement and the location of a 
Critical Point in the QCD phase diagram gave name to the series of 
CPOD conferences, which started 2004 with a workshop at the ECT* in Trento.} 
'Onset of Deconfinement' \cite{OD}, i.e., it seems that beam energies larger 
than $\sqrt{s_{NN}}\simeq (5-7)$~GeV are needed to
create in heavy ion collisions sufficiently dense matter 
so that the phase boundary to a high density deconfined state of matter
can possibly be crossed. Indications for the onset of deconfinement
now are also found in data for the elliptic flow parameter
$v_2$ measured by STAR in the BES at RHIC \cite{Shi}; for 
$\sqrt{s_{NN}} \le 11.5$~GeV $v_2$ for different hadron species no longer 
scales with the number of constituent quarks. At lower
beam energies the flow pattern seems to be controlled by purely hadronic 
degrees of freedom.

\begin{figure}
\begin{center}
\begin{minipage}{0.4\textwidth}
\begin{center}
\includegraphics[width=0.99\textwidth]{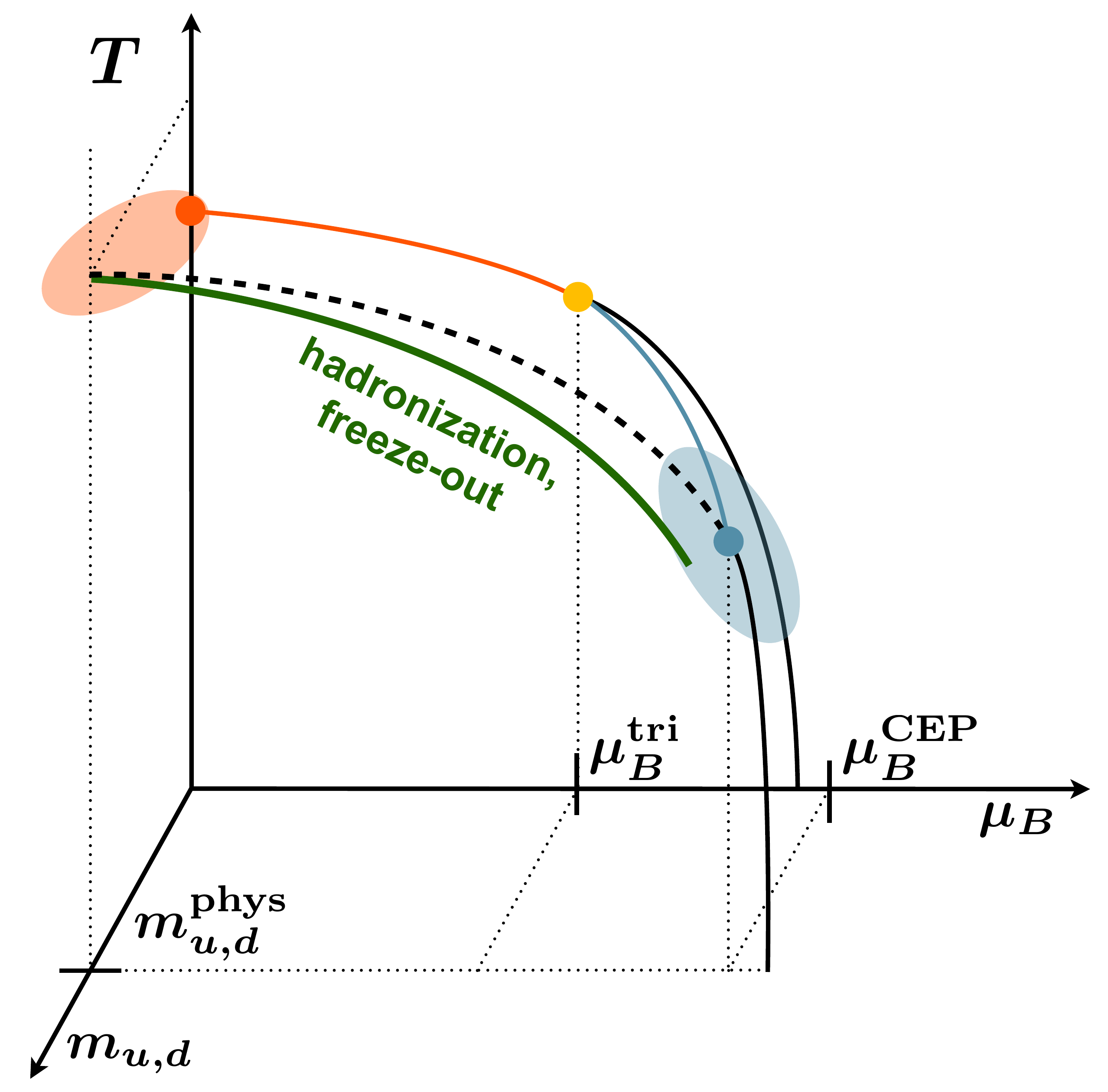}
\end{center}
\end{minipage}
\begin{minipage}{0.2\textwidth}
\end{minipage}
\begin{minipage}{0.5\textwidth}
\begin{center}
\includegraphics[width=0.99\textwidth]{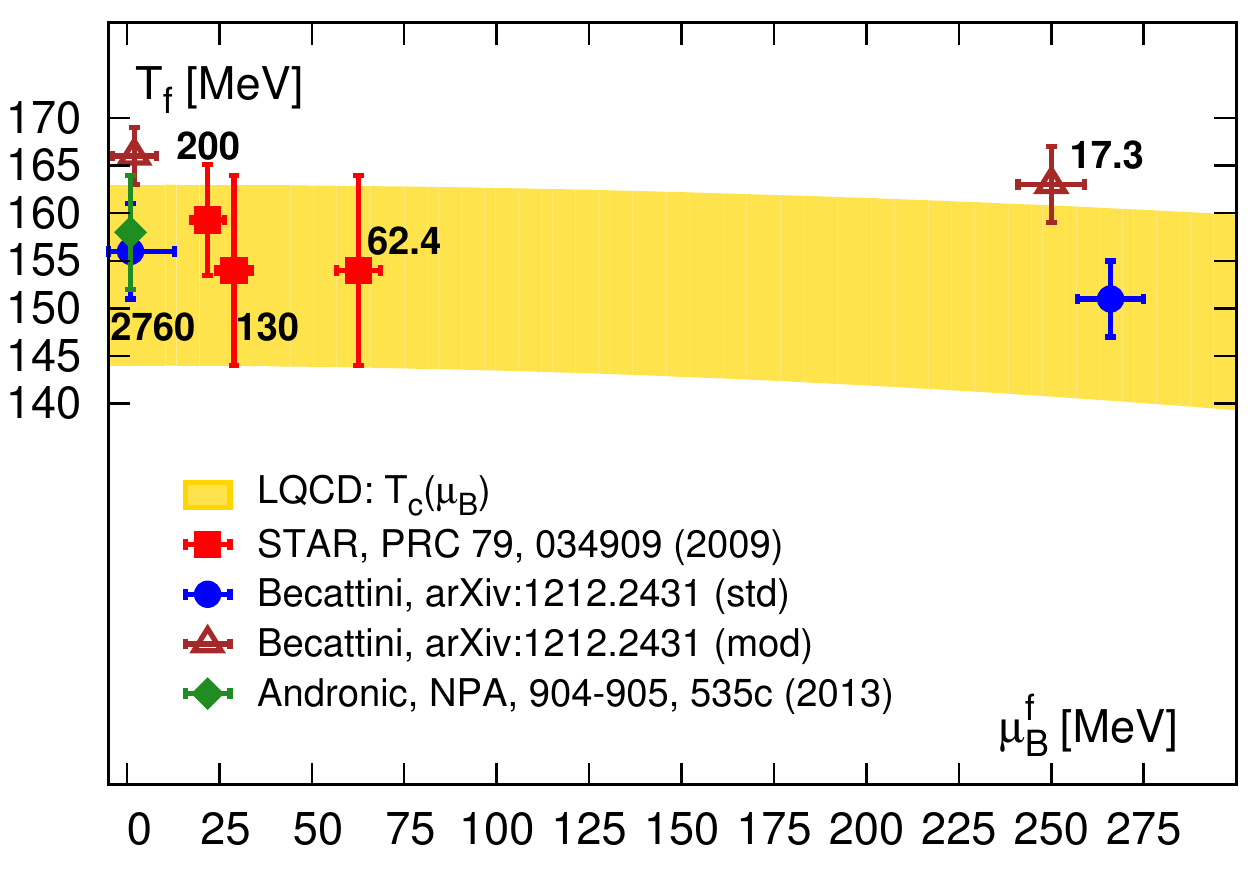}
\end{center}
\end{minipage}
\end{center}
\caption{{\bf Left:} A schematic phase diagram of QCD in the
$T$-$\mu_B$ and light quark mass space. It shows
the phase transition line at vanishing chemical potential
as well as a line of second order transitions, that will
exist, if a tri-critical point exists in the chiral limit at 
non-zero baryon chemical 
potential. The dotted line shows the crossover transition line 
of QCD at physical values of the quark mass that ends in a 
critical point. Also indicated is the freeze-out or 
hadronization line determined experimentally. Shaded areas 
indicate the critical region in the vicinity of the critical 
endpoint and the chiral transition at $\mu_B=0$, respectively.
{\bf Right:} The QCD transition temperature as function of the
baryon chemical determined from the maximum of the chiral 
susceptibility \cite{hotQCDTc}
and the scaling behavior of the second derivative of the chiral
condensate with respect to the chemical potential \cite{Kaczmarek}.
Data points show chemical freeze-out temperatures determined from
particle yields measured at the LHC \cite{Becattini,Andronic:2012dm}, 
the SPS \cite{Becattini} and in the BES at RHIC \cite{Abelev:2008ab}.
The difference between triangles and circles reflects the influence
of proton anti-proton annihilation processes, that take place after 
hadronization, on the determination of the freeze-out parameters
\cite{Becattini}.
\label{fig:phase}}
\end{figure}

While we gained confidence that a new form of matter
has been formed in heavy ion collisions, we
are less confident about experimental evidence for the location
of the phase boundary that separates the low and high temperature regime
and so far we also have no indications for the existence of a critical endpoint.
Lattice QCD calculations are ambiguous about the possible existence of a 
critical point \cite{Katz,Philipsen,Gavai} and
first results from the BES at RHIC are far from being conclusive 
\cite{STAR,Kumar:2012fb,Mitchell:2012mx}. The analysis of
net conserved charge fluctuations of hadrons may improve this situation. 
They carry information about critical behavior related to the occurrence
of a phase transition in the limit of vanishing quark mass, $m_q$, as 
well as to the existence of a critical endpoint at $\mu_B>0$ and $m_q>0$. 
However, in order for them to be a sensitive tool for detecting critical 
behavior in heavy ion experiments it clearly is necessary that the measurable
fluctuations are generated close to the QCD phase boundary (for an 
illustration see Fig.~\ref{fig:phase}~(left)). 

It generally is assumed that the fluctuations of conserved charges,
measured at RHIC or LHC, are generated at the time of chemical freeze-out.
Furthermore, it has been argued that the hadronization temperature, which 
one may identify with the QCD transition temperature, and the chemical 
freeze-out temperature, at which the abundances of different particle
species observed experimentally is fixed, differ only little 
\cite{BraunMunzinger:2003zz}.
This finds support through the recently performed reanalysis of 
particle yields at chemical freeze-out and the reconstructed yields 
at the time of hadronization \cite{Becattini}. This analysis
also shows that in particular the yields of protons
and anti-protons, which are important for the generation of net baryon
number fluctuations, may be distorted due to non-equilibrium, 
annihilation processes occurring between the time of hadronization and 
the freeze-out
of the bulk of particle species. This may also have consequences for the 
interpretation of experimentally observed charge fluctuations.

In order to use fluctuation observables in the search for critical
behavior it clearly is mandatory to understand at which time these fluctuations
are generated. I.e. are the fluctuations of thermal origin and, if so, what are 
the thermal conditions probed by
these observables? Do they correspond to a well defined point in the
QCD phase diagram? If so, how is this point related to the QCD transition
line? Answering these question also requires that we are clear about our
notion of the QCD transition temperature as well as the freeze-out
temperature.
We will address some of these questions, which have intensively 
been discussed at CPOD 2013, in the following.

\section{Pseudo-critical temperatures and freeze-out temperatures}

Searches for critical behavior in the BES
at RHIC and the
SPS to a large extent rely on the expectation that 
changes in correlation lengths that will arise in the vicinity of
a critical point give rise to {\it non-monotonic 
structures} in observables that are sensitive to fluctuations
generated at the time of hadronization or chemical freeze-out. 
This assumes that the line of chemical freeze-out 
parameters $(T_f,\mu_B^f)$ is well defined and closely follows the QCD 
transition line, which eventually may end in a true second order phase 
transition point. Of course, freeze-out itself is not an exact notion.
Not all particles will hadronize at the same time, some particle species
may fall out of equilibrium earlier than others and 
the decay of resonances at later times requires feed-down corrections.

Also the notion of a transition temperature in QCD is not exact. A true phase
transition with a unique critical temperature only exists in the chiral
limit, $m_q\rightarrow 0$. For $m_q>0$
one has several options to define pseudo-critical temperatures. However, also
other indicators for the transition are used, that have no direct
or only a weak link to critical behavior.
It thus is important to be clear about the notion of a pseudo-critical 
temperature and the way it is related to critical behavior in QCD as well 
as the chemical freeze-out temperature. 

{\bf Pseudo-critical temperatures.} 
In the absence of a true phase transition {\it
pseudo-critical} temperatures are introduced to characterize the crossover from the 
low temperature, chiral symmetry broken to the high temperature, chirally
symmetric phase of QCD. Pseudo-critical temperatures reflect properties of the 
non-analytic part of the free energy density. 
In the limit of vanishing explicit symmetry breaking, this non-analyticity 
shows up as a divergence in a sufficiently high order derivative of the
free energy density. In that limit all pseudo-critical 
temperatures approach the unique critical temperature. In order not to be
overwhelmed by regular contributions to the free energy density one clearly
should take care that the observables used to define a 
pseudo-critical temperature strongly couple to the singular part and, 
preferably, diverge at the critical point, i.e., in the limit 
of vanishing explicit symmetry breaking.

In QCD chiral symmetry is explicitly broken by the non-vanishing light
quark masses. The location of a peak in the disconnected part of the chiral 
susceptibility, $\chi_{disc} \sim \partial^2 \ln Z/\partial m_q^2$, 
provides a well defined pseudo-critical temperature.
The peak hight of $\chi_{disc}$
will increase with decreasing light quark mass $m_q$ and will
lead to a divergent susceptibility in the chiral limit. Its location 
defines the chiral phase transition temperature. 

In Fig.~\ref{fig:phase}~(right) we show results for the pseudo-critical 
temperature $T_c(\mu_B)$, determined from the position 
of a peak in $\chi_{disc}$ at  $\mu_B=0$ \cite{hotQCDTc}, and its leading 
order correction in $\mu_B/T$ \cite{Kaczmarek},
\begin{equation}
T_c(\mu_B) = \left( 154(9) - \kappa_B \left(\frac{\mu_B}{T}\right)^2 
\right)~{\rm MeV} \;\; {\rm with} \;\; \kappa_B=0.0066(7)
\; .
\label{Tc}
\end{equation} 
This is consistent with a determination of $T_c(\mu_B)$ that also is based
on an analysis of the chiral susceptibility \cite{WBTc,WB}. 

{\bf Strange quark number susceptibility and the Polyakov loop.}
In the chiral limit the QCD phase transition
is expected to belong to the universality class of three
dimensional $O(4)$ symmetric spin models
\footnote{This is not undisputed as the influence of the axial anomaly on the
chiral phase transition still is not well controlled.}. 
Derivatives of the order parameter
with respect to $T$ or $m_q$ become singular in that limit.
Divergences in energy like observables however only arise in third derivatives 
of the partition function with respect to temperature \cite{redlich}. I.e., 
second derivatives with respect to $T$, $\mu_B^2$ or $\mu_S^2$
stay finite even in the chiral limit. Quadratic fluctuations of conserved 
charges, which are only first derivatives of the QCD partition functions with 
respect to $\mu_B^2$ or $\mu_S^2$, show even less sensitivity for the singular 
structure of the QCD partition function and also their slope as function of 
temperature will stay finite in the chiral limit. In fact, extracting information
on critical behavior from quadratic quark number fluctuations would require
to become sensitive to a sub-leading temperature dependence, 
$\sim (T-T_c)^{1+|\alpha|}$. Extracting a transition temperature from
them thus is as difficult as extracting the transition temperature directly 
from the energy density of QCD. They are thus not well suited  for the 
definition of a pseudo-critical temperature.
Similar difficulties arise for the Polyakov loop expectation value ($P$) for 
which a direct relation to derivatives of the QCD partition function does not
exist, although $P$ is likely to couple to energy like observables. 

Nonetheless, the strange quark number 
susceptibility and the Polyakov loop expectation value have been used to
determine crossover temperatures. These observables lead to
estimates for the crossover temperature, which are  about 15~MeV
larger than the value extracted from the chiral susceptibility \cite{WBTc}.
This discrepancy may persist even in the chiral limit
as the singular contributions to these observables may be strongly
suppressed and overwhelmed by contributions from the regular part of 
the QCD partition function. The difference between these crossover temperatures
and the pseudo-critical temperature for the chiral transition led to speculations 
about a possible deconfinement transition being distinct from the chiral transition. 
We will discuss this in section 3.

{\bf Freeze-out temperatures.}
The line of freeze-out parameters is
commonly determined by comparing experimentally measured yields for various 
particle species with a statistical model calculation, i.e. the hadron 
resonance gas (HRG) model \cite{HRG}. This does provide a quite satisfactory
characterization of thermal conditions at the time of freeze-out
at the SPS \cite{Becattini} and in the BES at RHIC \cite{Abelev:2008ab}.
However, measurements of particle yields at the LHC made it obvious that this 
approach needs to be refined in order to deal with the
observed proton and anti-proton yields\cite{Becattini,Andronic:2012dm}.
As pointed out in Ref.~\cite{Becattini} the incorporation
of non-equilibrium processes arising in particular from proton anti-proton
annihilation and regeneration processes may improve thermal fits, but will 
lead to somewhat larger freeze-out temperatures \cite{Becattini} also at
lower beam energies. This
temperature may even agree with the hadronization temperature and thus should 
be close 
to the QCD transition temperature. Incorporating these effects in the 
determination of freeze-out parameters, however, requires the reconstruction 
of the latest state of hadronic chemical equilibrium using a transport model 
like UrQMD. This introduces additional model dependences in the determination
of freeze-out parameters. 

As shown by the triangles and circles in Fig.~\ref{fig:phase}~(right) 
the standard as well as the modified
approach to the determination of freeze-out parameters lead 
to temperatures consistent with the QCD transition temperature.
Although this is gratifying, it
also leaves us with a conceptual problem -- 
{\it as it seems that freeze-out happens indeed close to the QCD phase
boundary, one may wonder whether an uncorrelated HRG still captures 
correctly the strong
interactions that take place close to the QCD phase boundary.} At least
one should  verify that in this regime the HRG model still is a good 
approximation to QCD thermodynamics and, in particular, provides an 
appropriate characterization of the chemical composition of strongly 
interacting matter in the crossover region of the QCD transition. 

\section{The chiral transition and deconfinement}

Recent analyzes of quadratic charge fluctuations \cite{WBHRG,hotQCDHRG} have 
shown that these are 
indeed well described by the HRG for $T\lsim 160$~MeV. 
Moreover, strangeness fluctuations seem to be in agreement with HRG model 
calculations even at $T\simeq 170$~MeV. This led to speculations that 
strangeness may {\it deconfine} only well above the chiral transition 
\cite{Ratti}. 
We will discuss here that higher order cumulants do not provide any hints 
for such an assertion.  

{\bf Chiral transition.}
Cumulants of fluctuations of net conserved charges of QCD, i.e., net baryon 
number, electric charge and strangeness, are obtained from 
derivatives of the QCD partition function with respect to the corresponding
charge chemical potentials,
\begin{equation}
\chi_{ijk}^{BQS}(\mu_B,\mu_Q,\mu_S) = \frac{1}{VT^3}  
\frac{\partial^{i+j+k} \ln Z(V,T,\mu_B,\mu_Q,\mu_S)}{\partial(\mu_B/T)^i
\partial(\mu_Q/T)^j \partial(\mu_S/T)^k} \; .
\label{obs}
\end{equation}
Divergences arising from derivatives of the singular part of the QCD
partition function will show up prominently in higher order cumulants
$\chi_{ijk}^{BQS}$. 
These observables are thus sensitive to the 
occurrence of the chiral phase transition in QCD.
However, at 
vanishing $\mu_B$ only sixth order cumulants will diverge 
in the chiral limit and will provide an unambiguous 
signal for the chiral phase transition. For $\mu_B>0$ already third
order cumulants ($i+j+k=6$) will diverge. However, the leading
divergence is proportional to $(\kappa_B \mu_B/T)^3$, with $\kappa_B$ from
Eq.~\ref{Tc}. For $m_q>0$ the amplitude 
of the divergent term thus is parametrically suppressed and
may become important only at large $\mu_B$. In any case, 
an analysis of higher order cumulants of net charge fluctuations 
is needed in order to use charge fluctuations as a tool to study  the QCD 
chiral transition.

{\bf Deconfinement.}
The inflection point in the quadratic strange quark number susceptibility, 
$\chi_2^S\equiv \chi_{002}^{BQS}$, has been used to define a crossover 
temperature for the QCD transition
\cite{WBTc}.  As this temperature is about 15~MeV larger than
the pseudo-critical temperature extracted from the peak of the chiral 
susceptibility and as it
agrees with the temperature extracted from the slope of the Polyakov
loop, it has been speculated that deconfinement, in particular
deconfinement of strange quarks may happen in the chirally symmetric
phase \cite{Ratti}.
If correct, this  would also alter considerably our picture 
of strange charge fluctuations and their relation to chemical freeze-out.

\begin{figure}[t]
\begin{center}
  \includegraphics[width = 0.49\textwidth]{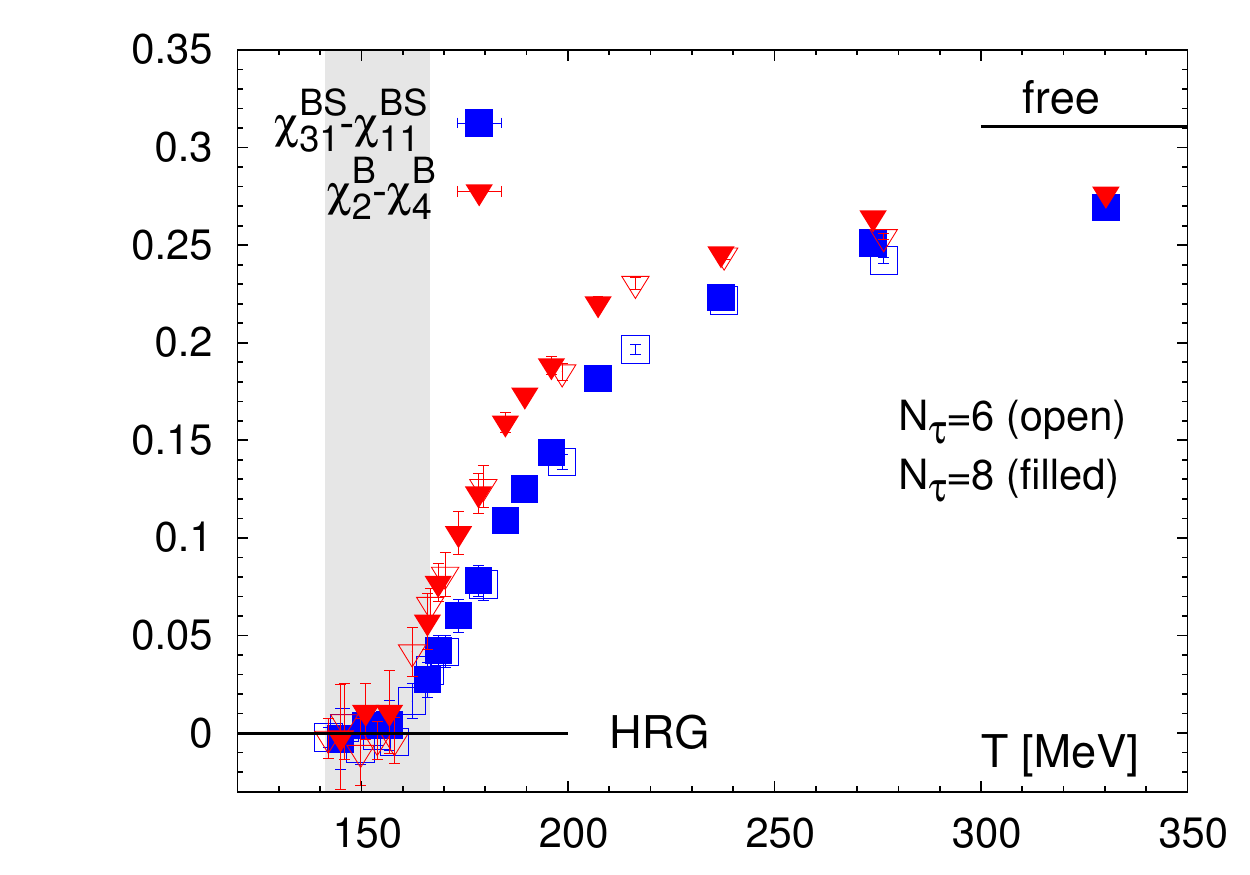}
  \includegraphics[width = 0.49\textwidth]{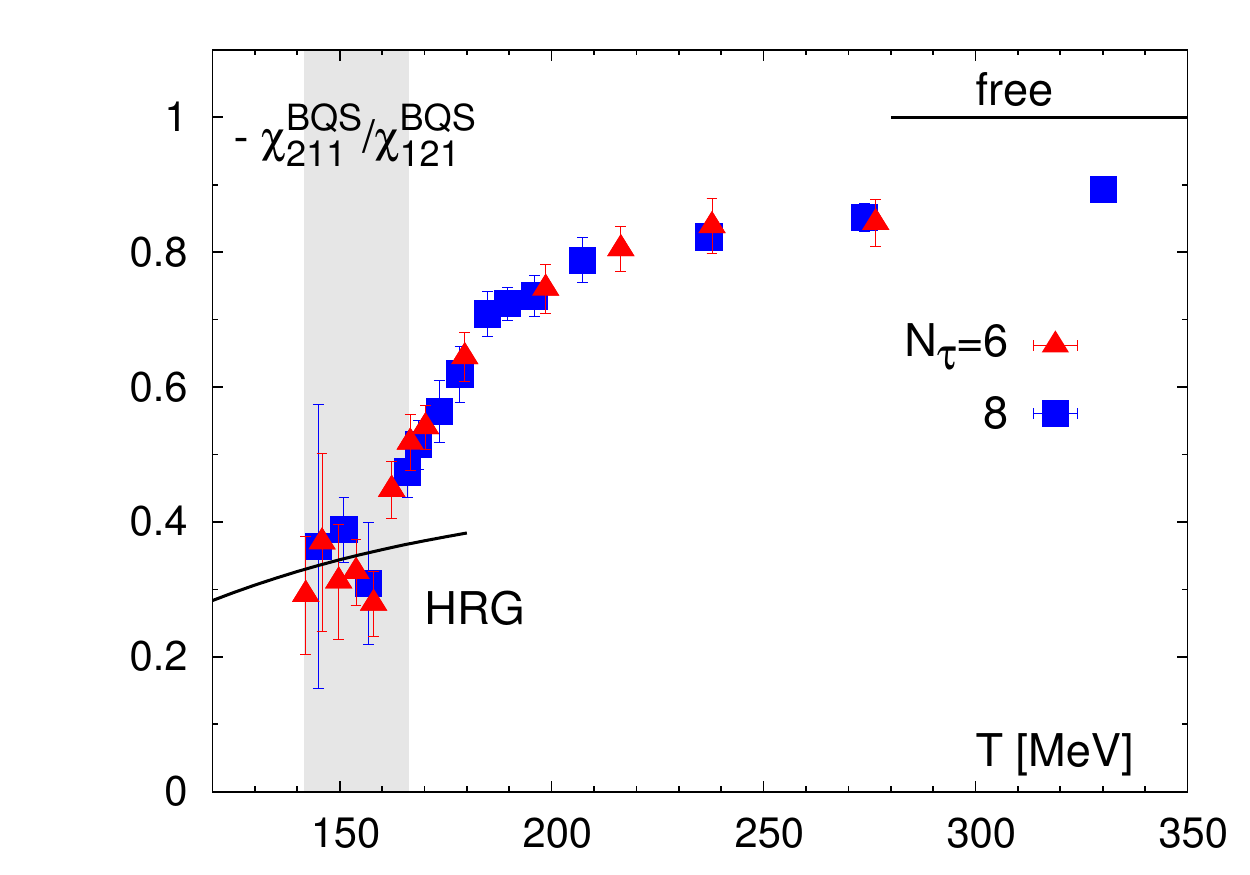}
\end{center}
\vspace*{-0.5cm}
  \caption{\label{fig:strange} {\bf Left:} The difference of $4^{th}$ and
$2^{nd}$ order cumulants combined in such a way that they vanish in an
uncorrelated hadron resonance gas (in Boltzmann approximation) 
\cite{BIBNL_strange}.
{\bf Right:} The ratio of two mixed cumulants that project onto the
quantum numbers of strange quarks and yield identical results in a free
strange quark gas.
}
\end{figure}

The analysis of fourth order cumulants of strangeness fluctuations provides
further insight into this question \cite{BIBNL_strange}. 
In Fig.~\ref{fig:strange}~(left) we show the difference between
the correlation of net strangeness fluctuations with first and third order
net baryon number fluctuations, respectively, i.e. 
$\chi_{31}^{BS} - \chi_{11}^{BS}$. In an uncorrelated gas of hadrons, i.e. 
the HRG, this difference vanishes. 
As can be seen from Fig.~\ref{fig:strange}~(left) this is approximately realized
only up to $T\simeq 160$~MeV. The same conclusion can be drawn from
the difference of second and fourth order net baryon number cumulants
$\chi_{2}^{B} - \chi_{4}^{B}$.
In a HRG also this quantity, which in addition includes information on baryons 
formed from light quarks only, vanishes. Fig.~\ref{fig:strange}~(left) strongly
suggests that the strange and non-strange baryon sector behave similar 
and that both differ significantly from an uncorrelated HRG for $T>160$~MeV.

Concurrent information on strangeness is also contained in the
ratio of mixed susceptibilities $\chi_{211}^{BQS}$ and $\chi_{121}^{BQS}$
shown in Fig.~\ref{fig:strange}~(right). At high temperature each of these 
susceptibilities singles out the same contribution to a free strange quark gas.
Their ratio thus will approach unity in the infinite temperature limit.
In a gas of uncorrelated hadrons these susceptibilities project
onto strange charged baryons, but they give different weight to the charge
one and two sectors. The latter is enhanced in $\chi_{121}^{BQS}$.
One thus expects the ratio $\chi_{211}^{BQS}/\chi_{121}^{BQS}$
to be less than unity in an HRG. As can be seen this is indeed the
case and the ratio is consistent with a HRG model calculation only for 
$T\lsim 160$~MeV.

Also the analysis of other fourth order cumulants \cite{Schmidt:2012kb}
leads to the conclusion that
the straightforward picture of an uncorrelated HRG breaks down
quite abruptly at
$T\simeq 160$~MeV, but provides a rather good description of hadron
thermodynamics and chemistry below this temperature. Also in the strange quark 
sector deviations from the HRG model rapidly become large for $T\ge 160$~MeV.
Strange hadrons thus are either dissolved or strongly modified above
$T=160$~MeV. This re-confirms that the pseudo-critical temperature
extracted from the chiral susceptibility is a good indicator for
critical behavior in QCD that also reflects deconfining features
of the QCD transition. 
At present there are no indications that
the liberation of light or strange quark degrees of freedom is delayed
and would happen at higher temperatures than the chiral crossover.

\section{Freeze-out and higher order cumulants of net charge fluctuations}

Higher order cumulants of net charge fluctuations become increasingly
sensitive to critical behavior. However, they 'only' influence the 
exponentially small tails of charge distributions. This makes it 
difficult to calculated them with high accuracy in lattice QCD as well as 
to measure them in heavy ion experiments. Nonetheless, they are appealing
because they allow to determine fundamental features 
of the QCD transition directly by comparing experimental data with 
a QCD calculation. 
E.g. one may examine ratios of cumulants that are
related to mean ($M$), variance ($\sigma^2$), skewness ($S$) or 
kurtosis ($\kappa$) of net charge distributions,
\begin{equation}
\frac{M_X}{\sigma_X^2} = \frac{\chi_1^X}{\chi_2^X}\;\; ,\;\;
\frac{S_X \sigma_X^3}{M_X} = \frac{\chi_3^X}{\chi_1^X}\;\; ,\;\;
\kappa_X \sigma_X^2 = \frac{\chi_4^X}{\chi_2^X}\;\; ,\;\;
X=B,\ Q,\ S \; .
\label{ratios}
\end{equation}

\begin{figure}[t]
\begin{center}
  \includegraphics[width = 0.89\textwidth]{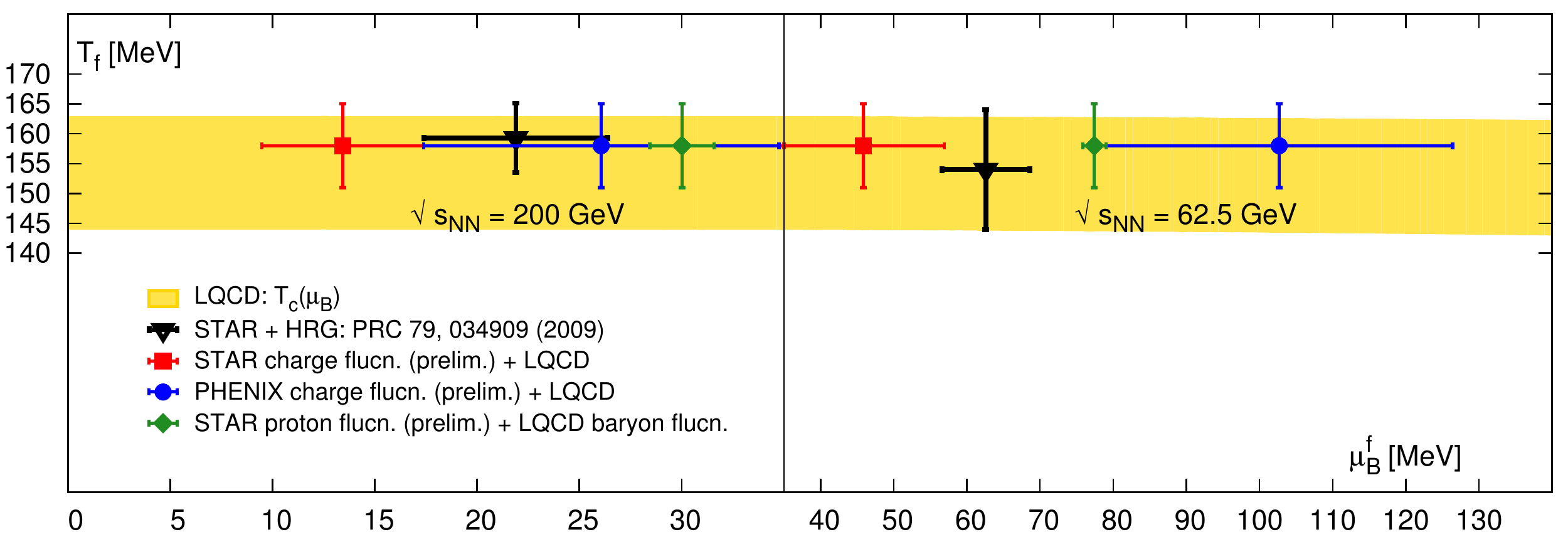}
\end{center}
\vspace*{-0.5cm}
  \caption{\label{fig:freeze} 
Freeze-out parameters determined from particle yields (triangle) \cite{Abelev:2008ab} 
and from net electric charge fluctuations (circles, squares) 
using the cumulants $\chi_1^Q/\chi_2^Q$
and $\chi_3^Q/\chi_1^Q$ to fix $\mu_B^f$ and $T_f$, respectively 
\cite{Swagato,Bazavov:2012vg}. Also shown is a
point obtained by comparing STAR data on net proton number fluctuations (diamond)
with lattice QCD calculations for net baryon number fluctuations 
\cite{Swagato}.
}
\end{figure}
The experimental determination of ratios of cumulants of net charge 
fluctuations itself is not a straightforward measurement. One needs
to control the influence of experimental cuts and efficiency corrections.
At lower beam energies one may worry whether the exact conservation of
charges in a small volume needs to be taken into account and 
one may need to correct for finite volume effects and volume fluctuations
\cite{Bzdak}.

In order to establish the measurement of fluctuations as a credible 
tool for the analysis of critical behavior in QCD a first step clearly
should be to verify that different cumulant ratios carry information
on thermal behavior that corresponds to a unique point in the QCD phase diagram. 
It thus is important to establish that thermal parameters $(T,\mu_B)$ and 
eventually also $(\mu_S,\mu_Q)$ can be extracted from cumulant ratios without 
making reference to model calculations. I.e. for consistency these parameters need 
to be determined by comparing QCD calculations with experiment. E.g. 
one may extract $(T_f,\mu_B^f)$ from any independent 
set of two ratios of cumulants. A convenient way to
do so is to use the ratios $\chi_1^X/\chi_2^X$ and $\chi_3^X/\chi_1^X$
\cite{Bazavov:2012vg}. 
Other ratios, involving also higher order cumulants, should then lead 
to a consistent determination of these freeze-out parameters.
In the BES at RHIC STAR and PHENIX have measured up to fourth order
cumulants of net electric charge fluctuations 
\cite{Kumar:2012fb,Mitchell:2012mx}
and STAR also presented results on net proton number fluctuations \cite{STAR}.
At present most of these data are preliminary. We use them here to test
whether a determination of $(T_f,\mu_B^f)$ from cumulant ratios as 
discussed above is feasible at all\footnote{The measured net proton cumulants 
may be considered
as a proxy for net baryon number fluctuations that can be calculated 
in lattice QCD. Of course, to some extent this assumes already 
that an uncorrelated gas of hadrons provides a good description of the 
thermodynamics at the time of freeze-out.
}.
Results of such an analysis are shown in Fig.~\ref{fig:freeze} \cite{Swagato}.
This necessarily preliminar analysis shows
that freeze-out temperature can be extracted from cumulant ratios and
are in reasonable agreement with the determination based on hadron yields. 
The current determination of 
freeze-out chemical potentials, however, also shows discrepancies
that become more obvious at lower beam energies. The discrepancies
between the preliminary data on electric charge fluctuations obtained by 
STAR and PHENIX clearly show that better control over the influence of cuts and
efficiency corrections is needed before a reliable determination
of freeze-out parameters from cumulant ratios can be performed. 
At present the analysis that led to Fig.~\ref{fig:freeze} should thus merely
be taken as a proof of principle.

\section{QCD critical endpoint}

\begin{figure}[t]
\begin{center}
  \includegraphics[width = 0.49\textwidth]{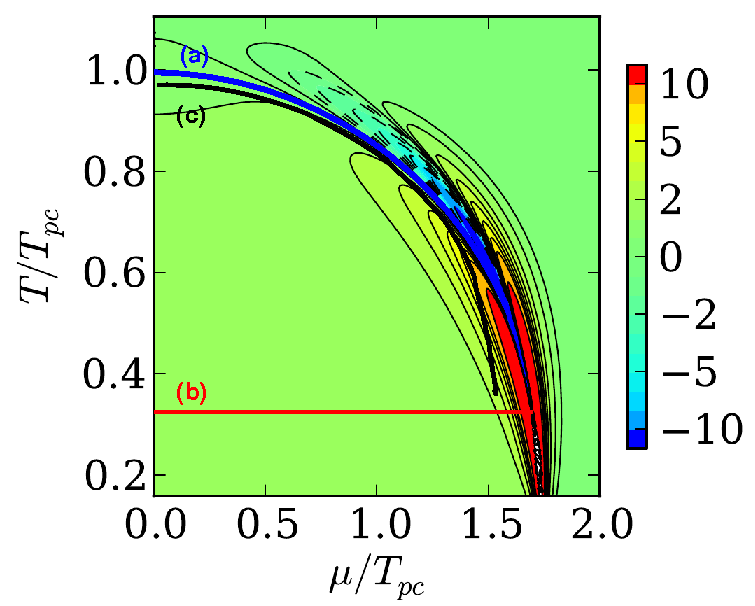}
\includegraphics[width = 0.49\textwidth]{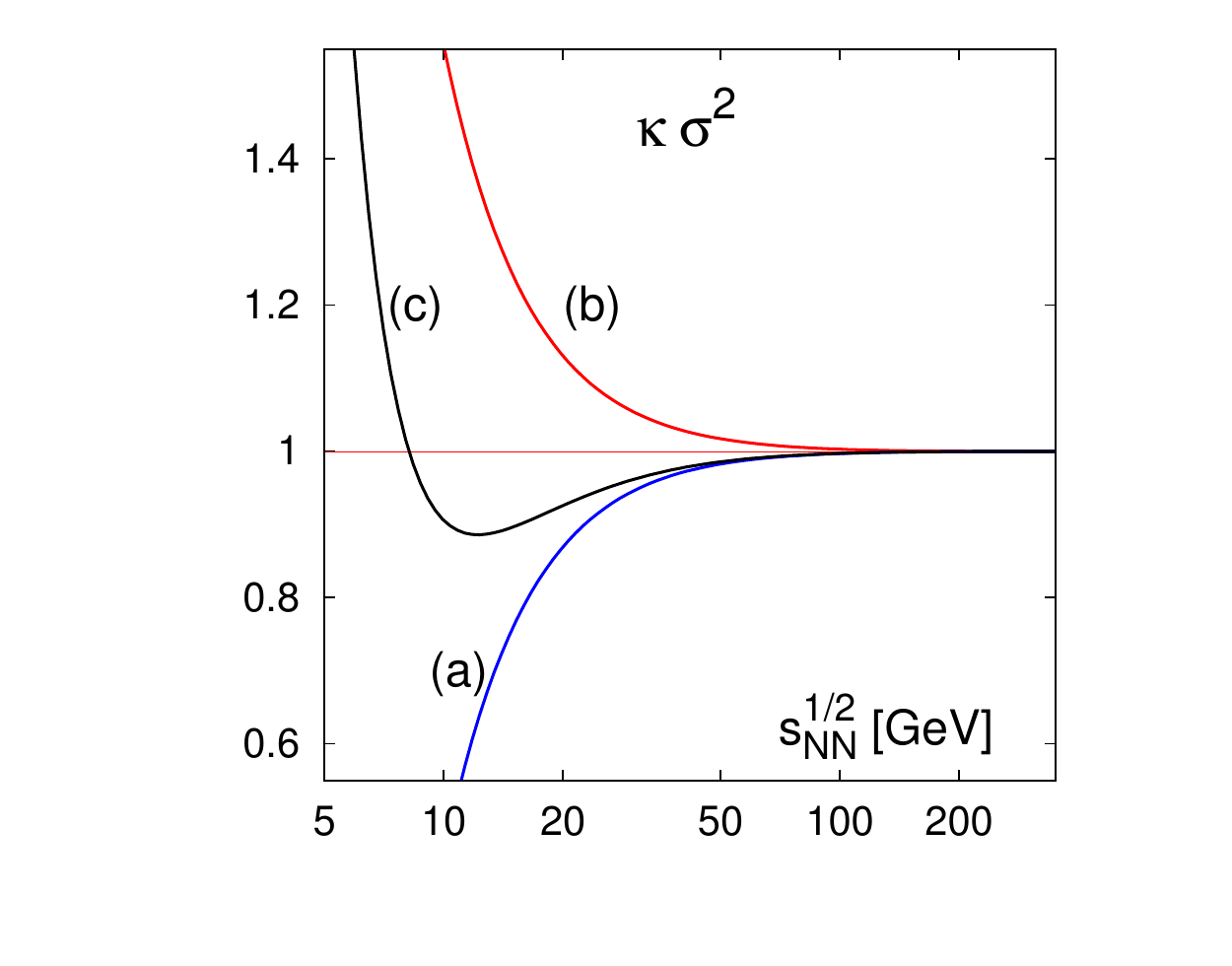}
\end{center}
\vspace*{-0.5cm}
  \caption{\label{fig:kurtosis}
Contour plot for the baryon number kurtosis times variance calculated
in the Polyakov-Quark-Meson model \cite{Skokov} (left) and a 
sketch for the beam energy dependence of this quantity for three
different paths in the $T$-$\mu_B$ plane.
}
\end{figure}

At least for small values of $\mu_B$, or for 
beam energies $\sqrt{s_{NN}}\ \gsim \ 20$~GeV, it seems that the measured 
net charge fluctuations are generated close to the QCD transition.   
If this remains to be the case at larger values of the chemical potential
one may indeed hope that fluctuations,
generated at the time of hadronization, carry information about a
nearby critical point and thus allow to detect its location in the
QCD phase diagram. The signature for the critical point encoded in the
beam energy dependence of e.g. $(\kappa \sigma^2)_B$ or $(\kappa \sigma^2)_Q$
nonetheless will depend strongly on the relative location of the freeze-out 
points and the critical point. This is easily illustrated in a model 
calculation. 

In Fig.~\ref{fig:kurtosis}~(left) we show a contour plot 
for net baryon number kurtosis times variance, $(\kappa \sigma^2)_B$,
calculated in a mean-field approximation of the quark meson model \cite{Skokov}.
The basic features seen in this figure arise from the global $O(4)$ 
symmetry that the model has in common with QCD and the existence of 
a critical point in this model. When increasing the
baryon chemical potential along the crossover line (path (a))
$(\kappa \sigma^2)_B$,
decreases, becomes negative and eventually will diverge to minus
infinity at the critical point. On the other hand, 
$(\kappa \sigma^2)_B$ stays positive when approaching the 
critical point at a lower temperature, in particular for $T=T_{CP}$
(path (b)). It will increase with increasing $\mu_B$ and eventually
will diverge to plus infinity when approaching the critical 
point\footnote{This feature is actually exploited when estimating the location
of the critical point from ratios of subsequent expansion coefficients
of the QCD partition function.}.
Along the hypothetical freeze-out line (path (c)) $(\kappa \sigma^2)_B$ 
may decrease with increasing $\mu_B$, eventually will have
a minimum and start to rise again.
In Fig.~\ref{fig:kurtosis}~(right) we show a sketch for
three possible scenarios for $(\kappa \sigma^2)_B$ as function of
$\sqrt{s_{NN}}$. A similar discussion, of course, also holds for
cumulants of net electric charge fluctuations.

It thus is obvious that the observable variation of $(\kappa \sigma^2)_B$ with 
beam energy will crucially depend on the location of the freeze-out line relative 
to the critical point. 

\section{Conclusions}

Higher order cumulants of net conserved charge fluctuations are 
sensitive to critical behavior in QCD. They provide clear-cut
signatures for the existence of a critical point in the QCD phase 
diagram. Whether the cumulants measured experimentally can 
convey this information crucially depends on the relative 
location of the QCD transition line and the freeze-out line
at which charge fluctuations are generated. In order to 
establish the measurement of higher order cumulants as a
signature for critical behavior in QCD we need to establish
that the measured fluctuations are indeed generated at a unique
point in the QCD phase diagram and that they reflect thermal conditions 
described by equilibrium QCD. The preliminary results on net proton
number and electric charge fluctuations from the BES at RHIC 
suggest that this will in the future become possible through a 
systematic comparison with cumulants calculated in lattice QCD.

\end{document}